# Microscopic description of a nonequilibrium system

E. N. Miranda
CRICYT – CONICET
5500 – Mendoza, Argentina
and
Departamento de Física
Universidad Nacional de San Luis
5700 – San Luis, Argentina

**Abstract:**
The empirical laws of chemical kinetics are studied from a microscopic point of view. An analysis based on elementary probability theory and combinatorics is enough to explain the kinetics law observed in experiments. Thus, an out of equilibrium system may be examined with tools available to a student who begins the study of statistical physics.

**PACS:** 5.40, 82.20

A first course in statistical mechanics is mostly devoted to the study of equilibrium systems, and little -or no- attention is paid to nonequilibrium ones. The reason is clear: the consideration of out of equilibrium problems are difficult and is left for advances courses. To change this situation, the kinetics of an elementary chemical reaction is examined in this article. It is a simple process that can be described with a basic knowledge of probability theory and combinatorics. Consequently, a beginner can understand it.

Since this is a journal devoted to physics students, some concepts of chemical physics are presented at the beginning and at the end of the article. In this way the paper is self-contained.

A central concept in chemistry is that of equilibrium [1][2]. Given the reactives $A$ and $B$, the products $C$ and $D$ come out from a chemical reaction. After some time, it is experimentally found that the concentrations of reactives and products reach a stationary state. These values are not independent but are related to one another through the law of mass action and an equilibrium constant $K_{eq}$. This law was inferred from experiments in 1867 by Guldberg and Waage and belongs to the background of every chemistry student. With the development of statistical mechanics, that law could be deduced from a microscopic model of matter. It was also possible to evaluate $K_{eq}$, related to the canonical partition functions of reactives and products. Today it is an exercise to reproduce those calculations. In this way the student realizes that a chemical law is not a fundamental one, but can be explained from equilibrium statistical mechanics. In this article we do the same task but with laws describing a chemical phenomenon out of equilibrium, i.e., the kinetics of elementary chemical reactions.

Let us consider a the following chemical reaction:

$$A + 2B \rightarrow C + D \tag{1}$$

It is said that this reaction is *elementary* if it really describes the actual molecular mechanism. This means that *1* molecule of $A$ interacts with *2* molecules of $B$ to produce *1* molecule of $C$ and *1* molecule of $D$. The numbers *1, 2, 1, 1* represent the molecularity of the reaction. It is not always so since the reaction (1) may not be elementary, but may be the net result of several steps. Each one of these steps is an elementary reaction –i.e. describes what really happens at the molecular level-, and eq. (1) is just the macroscopic result observed in the experiment. It is a task for the research chemist to decompose a given reaction into its elementary steps. In this paper we assume that eq. (1) represents an elementary reaction.

Experimentally it is found that the temporal evolution of the product concentrations [$C$] or [$D$] is given by:

$$\frac{d[C]}{dt} = \frac{d[D]}{dt} = v = k[A][B]^2 \tag{2}$$

As usual, [J] is the concentration of the chemical species J, v the reaction velocity and k is the so-called reaction constant. It should be remarked that eq. (2) is an empirical result, and our aim is to deduce this kinetic law from microscopic considerations.

Since (1) is an elementary reaction, at a microscopic level *1* molecule of A should combine with *2* molecules of B in a certain volume to generate the products. What happens if the concentration of B is doubled? In the same volume, there will be *4* molecules of B and *1* of A. But *2* molecules of B are enough to produce the reaction. There are several ways of choosing *2* molecules from *4* available; the total number of combinations is *4! / (2! 2!) = 6*. This result is surprising: a duplication in the concentration of B implies that the probability of the reaction -and consequently its speed- increases by a factor of *6* instead of *4* as shown by eq. (2). To solve this apparent inconsistency we should remember that a very large number of molecules are involved in the process.

If *b* is the initial number of B molecules, the number of different pairs that can be chosen is:

$$\binom{b}{2} = \frac{b(b-1)}{2} \tag{3}$$

It the number of B molecules is raised to *nb*, the relative increase in the number of pairs is given by:

$$\frac{nb(nb-1)}{2} \frac{2}{b(b-1)} = \frac{nb(nb-1)}{b(b-1)} \tag{4}$$

But we should remember that the typical number of molecules involved in a reaction is of the order of the Avogadro number, i.e., $b \approx N_A = 6 \times 10^{23}$. For such values one may write:

$$\frac{nb(nb-1)}{b(b-1)} \cong \frac{nb(nb)}{b(b)} = n^2 \tag{5}$$

Now everything fits. If the concentration is incremented n times, the probability that the reaction takes places grows $n^2$ times, and the reaction velocity increases in the same proportion. This implies that the velocity depends on the concentration as $v \propto [B]^2$. Since only one A molecule is needed for the reaction, the functional dependency for this reactive will be $v \propto [A]$. If a proportionality constant k is introduced, the reaction velocity comes out to be:

$$v = k[A][B]^2 \tag{6}$$

This is exactly the kinetic law found in experiments for reaction (1).

The reaction (1) may be generalized to:

$$xA + yB \rightarrow C + D \tag{7}$$

This reaction takes place if a group of *x* molecules of *A* interacts with a group of *y* molecules of *B*. If there are *a* molecules of *A* and *b* of B, there are many ways to choose those groups. A simple combinatorial analysis shows that the total number of combinations *W* is:

$$W = \binom{a}{x}\binom{b}{y} = \frac{a!}{x!(a-x)!}\frac{b!}{y!(b-y)!}$$
$$= \frac{a(a-1)....(a-x+1)}{x!}\frac{b!}{y!(b-y)!} \tag{8}$$

If the number of *A* molecules is increased to *na*, the total number of possible combinations changes to *W'*:

$$W' = \binom{na}{x}\binom{b}{y} = \frac{(na)!}{x!(na-x)!}\frac{(b)!}{y!(b-y)!}$$
$$= \frac{na(na-1)...(na-x+1)}{x!}\frac{(b)!}{y!(b-y)!} \tag{9}$$

The relative change in the number of possible combinations and consequently the relative change in the reaction speed is:

$$\lim_{\substack{a\to\infty\\b\to\infty}} \frac{W'}{W} = \lim_{\substack{a\to\infty\\b\to\infty}} \frac{na(na-1)....(na-x+1)}{a(a-1)...(a-x+1)} = n^x \tag{10}$$

The limit of $a\to\infty$ and $b\to\infty$ are taken due to the large number of molecules involved. Eq. (1) states that the number of ways for the reaction to take place increases $n^x$ times if the concentration of *A* increases *n* times. Therefore, the velocity depends on the concentration as $v \propto [A]^x$.

If the same analysis is performed changing the number of *B* molecules, increase *b* to *mb*, it is found that the relative increase of the possible combinations is $m^y$. Thus, the velocity dependency with the concentration is $v \propto [B]^y$. If a proportionality constant *k* is introduced, we end that the reaction velocity should be of the kind:

$$v = k[A]^x[B]^y \tag{11}$$

This is exactly the of kinetic law found experimentally for a reaction described by eq. (7)

To complete our analysis, we should describe what happens when equilibrium is reached. In that case, the inverse reaction has to be taken into account, i.e., it is possible that *1* molecule of *C* interacts with *1* molecule of *D* to produce *x* molecules of *A* and *y* molecules of *B*. The kinetic of the direct reaction is characterized by a constant *k* and the inverse one by *k'*. In symbols:

$$xA + yB \xrightarrow{k} C + D$$
$$C + D \xrightarrow{k'} xA + yB \qquad (13)$$

The velocity of the inverse reaction is:

$$v' = k'[C][D] \qquad (14)$$

In equilibrium the velocities of both reactions are equal and the concentration of each chemical species *J* reaches the value $[J]_{eq}$. Therefore, we may write:

$$\frac{[C]_{eq}[D]_{eq}}{[A]_{eq}^{x}[B]_{eq}^{y}} = \frac{k}{k'} = K_{eq} \qquad (15)$$

Eq. (15) is none but the law of mass action that was inferred from experiments in the XIX century but for us is an obvious consequence of the kinetic of a chemical reaction in equilibrium. For the special case when the intervening species are ideal gases, it can be shown [3][4] that $K_{eq}$ is given by:

$$K_{eq} = \frac{Z_C Z_D}{Z_A^x Z_B^y} \qquad (16)$$

$Z_J$ is the partition function of species *J*. The system is now in equilibrium and all the tools of statistical mechanics are applicable.

We have studied the kinetic of a chemical reaction that is an out-of-equilibrium process. From an elementary probabilistic analysis, we have inferred the functional dependence of the reaction velocity with the reactive concentrations. As has been shown the microscopic cause of kinetic laws are a direct consequence of the different possible combinations among molecules. One might say that chemical kinetics laws are an outcome of combinatorics. And once again, we can see that phenomenological laws can be explained from a microscopic description of matter and probabilistic considerations. This task is usually done for equilibrium systems; here it has been done for a nonequilibrium one that can be understood by a beginner in statistical physics.

**Acknowledgement:** The author thanks the National Research Council of Argentina (CONICET) for financial support.